\documentclass{sig-alternate-05-2015}

\def\Earg#1{\mathbb{E}\left[{#1}\right]}
\def\Esubarg#1#2{\mathbb{E}_{#1}\left[{#2}\right]}

\begin{document}

\setcopyright{acmcopyright}





%

\title{A Framework for Network A/B Test}
%
%
%
%
%

\numberofauthors{5} 
%
\author{
%
%
\alignauthor
Bai Jiang \titlenote{This author is a Ph.D. candidate in Statistics at Stanford University, who is now working at Yahoo Inc. as a research scientist intern.}\\
       \email{baijiang@yahoo-inc.com}
\alignauthor Xiaolin Shi\\
       \email{xishi@yahoo-inc.com}
\alignauthor
Hongwei Shang\\
       \email{shanghongwei@yahoo-inc.com}
 \and  
\alignauthor Zhigeng Geng\\
       \email{zgeng@yahoo-inc.com}
\alignauthor Alyssa Glass\titlenote{Corresponding author.}\\
       \email{alyssag@yahoo-inc.com}
}

\date{2 August 2016}


\maketitle
\begin{abstract}
A/B testing, also known as controlled experiment, bucket testing or splitting testing, has been widely used for evaluating a new feature, service or product in the data-driven decision processes of online websites. The goal of A/B testing is to estimate or test the difference between the treatment effects of the old and new variations. It is a well-studied two-sample comparison problem if each user's response is influenced by her treatment only. However, in many applications of A/B testing, especially those in HIVE of Yahoo and other social networks of Microsoft, Facebook, LinkedIn, Twitter and Google, users in the social networks influence their friends via underlying social interactions, and the conventional A/B testing methods fail to work. This paper considers the network A/B testing problem and provide a general framework consisting of five steps: data sampling, probabilistic model, parameter inference, computing average treatment effect and hypothesis test. The framework performs well for network A/B testing in simulation studies.
\end{abstract}

%
%
%
%

%
%



\section{Introduction}
A/B testing, also known as controlled experiment, bucket testing or splitting testing, has been widely used for evaluating a new feature, service or product in the data-driven decision processes of online websites, especially social networks of Yahoo, Microsoft, Facebook, LinkedIn, Twitter and Google \cite{kohavi2009controlled, kohavi2012trustworthy, kohavi2013online, kohavi2014seven}. The goal of A/B testing is to estimate or test the average treatment effect (ATE), which is defined as the difference between the treatment effects of the old and new variations.


It is commonly assumed in the past industrial practice that the response of each user relies on her treatment only, no matter what treatments other users receive and what response other users give. Such an independence condition is known as Stable Unit Treatment Value Assumption (SUTVA) \cite{rubin1978bayesian, rubin1980comment, rubin1990comment}. Another common assumption is that the responses of users who receive the same treatment are identically distributed. Under these two assumptions, one can view groups A and B as two independent and identically distributed (i.i.d.) samples and solves the A/B testing problem as a two-sample comparison problem, which is well-studied.

However, the SUTVA does not hold in many important applications of A/B testing in social networks. the SUTVA for the classic A/B testing problem is violated in the user network where people interact with each other.  

The responses of users in an A/B test in a social network is usually a mixture of \textit{treatment effect}, \textit{network effect} and \textit{spill-over effect}. The \textit{network effect}, also known as social interactions, peer influence or social interference \cite{eckles2014design}, results from the correlation of a user's behavior and her neighbors'. Since users exchange information and interact with their neighborhood in a social network, the response of each user is influenced by those of her neighbors and vice versa \cite{toulis2013estimation}. Furthermore, people who share similar interests and preferences tend to cluster together in a community and exhibit similar response to a new feature, new service or new product. In some scenarios, the new product produces effect only if both a user and a couple of her neighbors are exposed to it, and thus the network effect must be considered and included in the A/B test. For example, Yahoo's mobile app product HIVE provides a community platform where users could ask and answer questions to seek information and support. Most A/B tests on HIVE require intraction among a few users.  The \textit{spill-over effect} happens if some users assigned to group A interact with their neighbors assigned to group B so that the treatment effect of A may spill over to group B and vice versa.

Recently A/B testing in social networks has gained sharpened focus (see \cite{backstrom2011network, ugander2013graph, gui2015network} among others). The issues in the network A/B testing problem have been addressed from different aspects. Backstorm and Kleinberg \cite{backstrom2011network} described a random-walk-based sampling method for producing samples of users (nodes in the network) that are internally well-connected but also approximately uniform over the population. Ugander et. al. \cite{ugander2013graph} show in a simplified setting how graph cluster randomization produce an unbiased estimate for ATE with asymptotically small variance. Gui et. al. \cite{gui2015network} proposed several linear additive models and conducted experiments on data sets from real social networks.

This paper builds a general framework for network A/B testing, which consists of 5 steps: data sampling, probabilistic model, parameter inference, ATE computation, and hypothesis test. Such a framework is compatible to many existing studies on the network A/B testing problem and enables a more comprehensive solution. In particular, we proposed a bootstrapping method for the hypothesis test $\text{ATE}>0$, which is lacked by previous studies.

Section 2 formally describe the network A/B testing problem. Section 3 presents five steps in the framework: data sampling, probabilistic model, parameter inference, ATE computation, and hypothesis test. Section 4 show simulation results, which suggests the potential of our methods for network A/B testing on real data sets.

\section{From A/B Testing to Network A/B Testing}
Denote by $V$ the set of users, by $Z_i=0$ or $1$ whether user $i \in V$ is assigned to group A or B, and by $Y_i$ its response. The average treatment effect (ATE) is defined by (\ref{eqn: ATE}).
\begin{equation}
\text{ATE} \triangleq \Earg{\frac{1}{|V|}\sum_{i \in V}Y_i|\mathbf{Z}=\mathbf{1}} - \Earg{\frac{1}{|V|}\sum_{i \in V}Y_i|\mathbf{Z}=\mathbf{0}} \label{eqn: ATE}
\end{equation}
If two group of users $\{Y_i: i \in A\}$ and $\{Y_i: i \in B\}$ follow the distributions $Y|Z=0$ and $Y|Z=1$ i.i.d. respectively, then
\begin{align*}
\Earg{Y|Z=0} &= \Earg{\frac{1}{|V|}\sum_{i \in V}Y_i|\mathbf{Z}=\mathbf{0}},\\
\Earg{Y|Z=1} &= \Earg{\frac{1}{|V|}\sum_{i \in V}Y_i|\mathbf{Z}=\mathbf{1}},
\end{align*}
and thus the ATE (\ref{eqn: ATE}) is simplified as 
$$\text{ATE} = \mathbb{E}[Y|Z=1] - \mathbb{E}[Y|Z=0].$$
From the strong law of large number it follows that
\begin{align*}
\frac{1}{|A|} \sum_{i \in A} Y_i  &\overset{a.s.}{\to} \Earg{Y|Z=0}\\
\frac{1}{|B|}\sum_{i \in B} Y_i &\overset{a.s.}{\to}  \Earg{Y|Z=1}
\end{align*}
as more and more users are involved in the experiment, which further implying that the difference between the average responses of group A and B is an consistent estimate for ATE
$$\frac{1}{|B|}\sum_{i \in B} Y_i - \frac{1}{|A|} \sum_{i \in A} Y_i  \overset{a.s.}{\to} \text{ATE}$$
The significance of the point estimate for ATE is usually evaluated by a statistical hypothesis test and indexed by a p-value. To statistically compare two i.i.d. samples $\{Y_i : i \in A\}$ and $\{Y_i : i \in B\}$, it is natural to apply Student's t test or Fisher's exact test.

\begin{figure}[h!]
\centering
\includegraphics[width=0.2\textwidth]{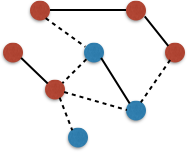}
\caption{Treatment effect, network effect, spill-over effect in the network A/B testing.} \label{fig: spill-over effect}
\end{figure}

However, the i.i.d. assumption (or SUTVA) does not hold in the setting of network A/B testing, the responses of users in a social network is a mixture of \textit{treatment effect}, \textit{network effect} and \textit{spill-over effect}. Figure \ref{fig: spill-over effect} illustrates a small social network of 8 users (nodes) and 9 friendship relationships (edges). Five users belong to group A (red), while the other three users belong to group B (blue). Solid lines present the \textit{network effect}, namely the peer-to-peer influence within either group A or B. Dashed lines present the \textit{spill-over effect}, namely the peer-to-peer influence across group A and B.

Denote by $G(V,E)$ the underlying social network, where $V$ is the set of nodes, and $E$ is the set of edges. Let $\mathbf{Z} = \{Z_i: i \in V\}$ the vector of assignments of each user belonging to either group A or B, and $\mathbf{Y} = \{Y_i: i \in V\}$ the vector of responses of each user. On the social network $G(V,E)$, it is a challenging task to estimate the ATE (\ref{eqn: ATE}) given the data of users' response $\mathbf{Y}$ being ``contaminated" by both \textit{network effect} and \textit{spill-over effect} via social interactions $E$. The difference between two sample averages is no longer an consistent estimate for the ATE (\ref{eqn: ATE}), i.e.
\begin{align*}
\frac{1}{|A|} \sum_{i \in A} Y_i  &\overset{a.s.}{\not \to} \Earg{\frac{1}{|V|}\sum_{i \in V}Y_i|\mathbf{Z}=\mathbf{0}}\\
\frac{1}{|B|}\sum_{i \in B} Y_i &\overset{a.s.}{\not \to}  \Earg{\frac{1}{|V|}\sum_{i \in V}Y_i|\mathbf{Z}=\mathbf{1}}\\
\frac{1}{|B|}\sum_{i \in B} Y_i - \frac{1}{|A|} \sum_{i \in A} Y_i  &\overset{a.s.}{\not \to} \text{ATE}
\end{align*}
as more and more users are involved. Moreover, neither Student's t test nor Fisher's exact test is applicable for the comparison of two non-i.i.d. and dependent samples. 

\section{A Framework for Network A/B Testing}
This paper aims to provide a general framework for network A/B testing, which consists of 5 steps:
\begin{itemize}
\item data sampling,
\item probabilistic model, 
\item parameter inference,
\item ATE computation, and 
\item hypothesis test.
\end{itemize}
Such a framework is compatible to many existing studies on the network A/B testing problem and enables a more comprehensive solution.

\subsection{Data Sampling}
As before, we denote by $G(V,E)$ the underlying social (sub-)network of users involved in the A/B test, by $\mathbf{Z}=\{Z_i: i \in V\}$ the vector of assignment, and by $\mathbf{Y}=\{Z_i: i \in V\}$ the vector of response. The step of data sampling is to sample a triplet $(G, \mathbf{Z}, \mathbf{Y})$ for A/B test in real or simulated a social networks.

\begin{figure}[h!]
\centering
\includegraphics[width=0.45\textwidth]{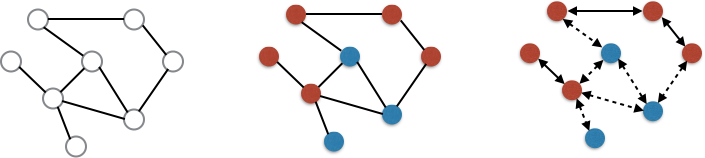}
\caption{Data sampling process.} \label{fig: data sampling}
\end{figure}

As we show in Figure \ref{fig: data sampling}, a typical data sampling process for network A/B test include: generating a subnetwork $G$ in the simulation study or select a subnetwork $G$ from the whole social network, assigning $Z_i = 0/1$ to each user $i$ by some rule, letting the users interact and collecting their response $Y_i$.

Many previous studies primarily focus on sampling $G(V,E)$ and/or $\mathbf{Z}$. For example, Backstorm and Kleinberg \cite{backstrom2011network} discuss a few random-walk-based methods for sampling $V$. Ugander et. al. \cite{ugander2013graph} select users if and only if they and (part of) their neighborhood are assigned to the same group. Gui et. al. \cite{gui2015network} partition the whole social network into multiple clusters, treat each cluster as a unit and randomize at the cluster
level, so that all the users in the same cluster are assigned to the same group.

Researchers also considered both group-level and user-level assignments of treatment/control. The group-level assignment allows no or little peer-to-peer interaction across groups \cite{sobel2006randomized, hudgens2012toward, rosenbaum2012interference, tchetgen2012causal, gui2015network}. The user-level assignment allows non-trivial peer-to-peer interaction between any pair of users no matter which groups they belong to \cite{toulis2013estimation, manski2013identification, eckles2014design, ugander2013graph}.

In our simulation study, we sample $G(V,E)$ by the small-world algorithm (i.e. Watts-Strogatz algorithm \cite{watts1998collective}). Other options for random graph model include geometric method \cite{penrose2003random}, Kronocker method \cite{leskovec2005realistic}, Barab{\'a}siand-Albert method \cite{albert2002statistical} and so on. $\mathbf{Z}$ is assigned at the user level. Specifically, $Z_i \sim \text{Bernoulli}(p)$ i.i.d. where $p$ is the proportion of users who belong to the treatment group B.

\subsection{Probabilistic Model}
Markov Random Field is the standard model for the social network in which random variables $\mathbf{Y}=\{Y_i: i \in V\}$ have a jointly distribution in an undirected graph $G(V,E)$ given the treatment/control assignment $\mathbf{Z}$. Denote by $\theta$ the parameter of the model, and by $P_\theta(\mathbf{Y}|\mathbf{Z};G)$ the joint distribution of $\mathbf{Y}$ given $\mathbf{Z}$ on social network $G(V,E)$. It is worth noting the Markov property in these models: $Y_i$ relies on other variables through its neighborhood $\{Y_j: j \in N(i) \}$ only. Formally speaking, let $n(i) = \{j \in V: (j,i) \in E\}$ be the neighborhood of user $i$, then
\begin{align*}
&\quad P_\theta\left(Y_i | \{Y_j: j \in V, j \ne i\}, \mathbf{Z}; G\right)\\
&= P_\theta\left(Y_i | Z_i, \{(Z_j,Y_j: j \in n(i) \}; G\right).
\end{align*}

One special case of Markov Random Field is Gaussian Graphical Model, in which $\mathbf{Y}$ follows a multivariate Gaussian distribution from the global view and each $Y_i$ follows an univariate Gaussian distribution conditional on its neighborhood from the local view. Examples are the additive linear models in \cite{gui2015network} like (\ref{eqn: GGM}).
\begin{align}
& \quad Y_i | Z_i, \{(Z_j,Y_j: j \in n(i) \}; G \nonumber\\
&= \underbrace{\alpha_0 + (\alpha_1-\alpha_0) Z_i}_\text{treatment effect} + \underbrace{\beta \sum_{j\in n(i)}Z_j + \gamma \frac{\sum_{j \in n(i)}Y_j}{|n(i)|}}_\text{network \& spill-over effects} + \epsilon_i \label{eqn: GGM}
\end{align}
where $\epsilon_i \sim \mathcal{N}(0,\sigma^2)$ and the parameters $\theta = (\alpha_0, \alpha_1, \beta, \gamma, \sigma)$. Here $\alpha_0, \alpha_1$ characterize the intensity of treatment effect of variations A and B, respectively. $\beta$ and $\gamma$ together characterize the intensity of network effect and spill-over effect.

Another special case is Ising/logistic Model, in which $Y_i \in \{-1,+1\}$ to present whether user $i$ gives negative or positive response. In this paper, we propose a variant of Ising model (\ref{eqn: Ising}) to fit negative/positive responses in the network A/B testing.
\begin{align}
P_\theta(\mathbf{Y}|\mathbf{Z};G) &\propto \text{exp} \underbrace{\left(\alpha_0 \sum_{i \in A} Y_i  + \alpha_1 \sum_{i \in B} Y_i \right.}_\text{treatment effect} \nonumber\\
&\qquad+ \underbrace{\beta_0 \sum_{(i,j)\in E; i,j \in A} Y_iY_j + \beta_1\sum_{(i,j)\in E; i,j \in B} Y_iY_j}_\text{network effect} \nonumber\\
&\qquad+ \underbrace{\left.\gamma \sum_{(i,j)\in E; i \in A, j \in B} Y_iY_j \right)}_\text{spill-over effect} \label{eqn: Ising}
\end{align}
where the parameters $\theta = (\alpha_0, \alpha_1, \beta_0, \beta_1, \gamma)$. Here $\alpha_0, \alpha_1$ characterize the intensity of treatment effect of variations A and B, respectively; $\beta_0,\beta_1$ for network effect among groups A and B, respectively; and $\gamma$ for spill-over effect across groups A and B.

From the local view, each $Y_i$ follows a logistic model (\ref{eqn: logistic A}) or (\ref{eqn: logistic B}) conditional on its neighborhood and its group assignment $Z_i$. The logistic function $\text{logistic}(x) = 1/(1+\exp(-x))$.
\begin{align}
\quad & P_\theta\left(y_i = +1| Z_i = 0, \{(Z_j,Y_j: j \in n(i)\}; G\right) \label{eqn: logistic A}\\
 = & \text{logistic}\left(\underbrace{2\alpha_0}_\text{treatment effect} + \underbrace{2\beta_0 \sum_{j \in n(i) \cap A } y_j }_\text{network effect} + \underbrace{2\gamma \sum_{j \in n(i) \cap B } y_j}_\text{spill-over effect}\right) \nonumber
 \end{align}
\begin{align}
\quad & P_\theta\left(y_i = +1| Z_i = 1, \{(Z_j,Y_j): j \in n(i)\}; G\right) \label{eqn: logistic B}\\
 = & \text{logistic}\left(\underbrace{2\alpha_1}_\text{treatment effect} + \underbrace{2\beta_1 \sum_{j \in n(i) \cap B } y_j }_\text{network effect} + \underbrace{2\gamma \sum_{j \in n(i) \cap A } y_j}_\text{spill-over effect}\right) \nonumber
 \end{align}
 
\subsection{Parameter Inference}
This subsection presents how to infer parameters $\theta$ by fitting $K$ triplets $(G^{(k)},\mathbf{Z}^{(k)}, \mathbf{Y}^{(k)})$ to the model $P_\theta(\mathbf{Y}|\mathbf{Z}; G)$.

It is challenging to yield the Maximum Likelihood Estimate (MLE) by solving the optimization problem \ref{eqn: MLE}).
\begin{equation}
\max_\theta \prod_{k=1}^K p_\theta(\mathbf{Y}^{(k)}|\mathbf{Z}^{(k)};G^{(k)}) \label{eqn: MLE}
\end{equation}
The difficulties are that the normalizing constant of the probability function $P_\theta(\mathbf{Y}|\mathbf{Z}; G)$ is usually unknown in Markov Random Field, and that the probability function $P_\theta$ does depend on the network structure $G$, which varies for different triplets $(G^{(k)},\mathbf{Z}^{(k)}, \mathbf{Y}^{(k)})$.

A remedy is to replace the likelihood function $P_\theta(\mathbf{Y}|\mathbf{Z}; G)$ in the objective function of (\ref{eqn: MLE}) with a pseudo-likelihood function $\prod_{i \in V} P_\theta(Y_i| Z_i,\{ Z_j,Y_j: j \in n(i)\};G)$, which is the product of conditional probabilities of $Y_i$. It results in the Maximum Pseudo-likelihood Estimate (MPLE), which is given by (\ref{eqn: MPLE}).
\begin{equation}
\max_\theta \prod_{k=1}^K \underbrace{\prod_{i \in V^{(k)}} p_\theta(Y_i^{(k)}| Z_i^{(k)},\{ Z_j^{(k)},Y_j^{(k)}\}_{j \in n(i)};G^{(k)})}_\text{pseudo-likelihood}. \label{eqn: MPLE}
\end{equation}
The MPLE is doable since the conditional probabilities of $Y_i$ are available in either (\ref{eqn: GGM}) for Gaussian Graphical Model or (\ref{eqn: logistic A}), (\ref{eqn: logistic B}) for Ising/logsitic Model. Moreover, solving MPLE is equivalent to fit linear or logistic regression just like all $\sum_{k=1}^K |V^{(k)}|$ users being independent samples.

\subsection{Computing Average Treatment Estimate}
Proceed to compute the ATE (\ref{eqn: ATE}) given $\theta$. It involves the computation of two terms, which are in essence complicated integrals with respect to two joint probability functions $P_\theta(\mathbf{Y}|\mathbf{Z}=\mathbf{1};G)$ and $P_\theta(\mathbf{Y}|\mathbf{Z}=\mathbf{0};G)$. Although the normalizing constants in the two joint probability functions are known for neither Gaussian Graphical Model nor Ising/logistic Model, we can still approximately compute the two terms by Gibbs sampling with the conditional probabilities like (\ref{eqn: GGM}), (\ref{eqn: logistic A}), (\ref{eqn: logistic B}).

\subsection{Hypothesis Test}
\begin{figure}[h!]
\centering
\includegraphics[width=0.48\textwidth]{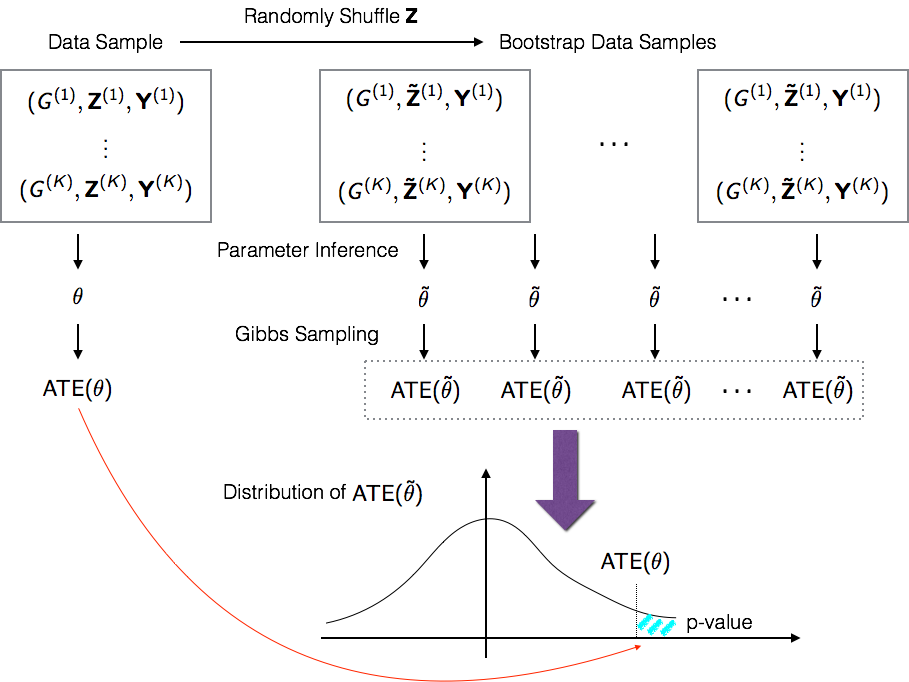}
\caption{The scheme of the bootstrapping method for p-value construction.} \label{fig: bootstrap scheme}
\end{figure}

Along with a point estimate for $\text{ATE}(\theta)$, a p-value is desired to indicate the confidence of the estimate. Unfortunately, to the best of our knowledge, there exists no existing method for hypothesis test $\text{ATE}>0$ so far. In this paper, we propose to construct a p-value by the bootstrapping method. Specifically,
\begin{enumerate}
\item Use data
$$(G^{(1)},\mathbf{Z}^{(1)},\mathbf{Y}^{(1)}), \dots, (G^{(K)},\mathbf{Z}^{(K)},\mathbf{Y}^{(K)})$$ to fit the model $P_\theta(\mathbf{Y}|\mathbf{Z};G)$ and compute $\text{ATE}(\theta)$.
\item Randomly shuffle $\mathbf{Z}^{(k)}$ as $\mathbf{\tilde{Z}}^{(k)}$, use bootstrapping data sample
$$(G^{(1)},\mathbf{\tilde{Z}}^{(1)},\mathbf{Y}^{(1)}), \dots, (G^{(K)},\mathbf{\tilde{Z}}^{(K)},\mathbf{Y}^{(K)})$$
to fit the model $p_\theta(\mathbf{y}|\mathbf{z};G)$ and compute $\text{ATE}(\tilde{\theta})$.
\item Repeat step 2 many times, and compare $\text{ATE}(\theta)$ to the bootstrapping distribution of $\text{ATE}(\tilde{\theta})$.
\end{enumerate}
If variants A and B do have significant treatment effect, $\text{ATE}(\theta)$ is expected to be extremely large compared to the bootstrapping distribution of $\text{ATE}(\tilde{\theta})$, which results from randomly pairing the response $Y_i$ and $Z_i$ on the network $G$. The p-value is the area beyond $\text{ATE}(\theta)$ under the curve of the bootstrapping distribution. Figure \ref{fig: bootstrap scheme} illustrates the whole scheme of our bootstrapping method.
 
It is also worth noting that to test $\text{ATE}(\theta)>0$ is equivalent to test $\alpha_1 > \alpha_0$ in the Ising/logistic model (\ref{eqn: Ising}), if $\beta_0 \approx \beta_1$. Indeed, assigning all users to group A, i.e. $\mathbf{Z}=\mathbf{0}$, in the model (\ref{eqn: Ising}) yields
\begin{equation}
P_\theta(\mathbf{Y}|\mathbf{Z}=\mathbf{0}; G) \propto \exp \left(\alpha_0 \sum_{i \in V} Y_i + \beta_0 \sum_{j \in n(i)} Y_iY_j\right). \label{eqn: Ising A}
\end{equation}
(\ref{eqn: Ising A}) is an exponential family. $\Esubarg{\theta}{\sum_{i \in V} Y_i|\mathbf{Z}=\mathbf{0};G}$, the expectation of the sufficient statistic $\sum_{i \in V} Y_i$, is increasing with respect with its coefficient $\alpha_0$. Similarly, assigning all users to group B, i.e. $\mathbf{Z}=\mathbf{1}$, in the model (\ref{eqn: Ising}) yields
\begin{equation}
P_\theta(\mathbf{Y}|\mathbf{Z}=\mathbf{1}; G) \propto \exp \left(\alpha_1 \sum_{i \in V} Y_i + \beta_1 \sum_{j \in n(i)} Y_iY_j\right). \label{eqn: Ising B}
\end{equation}
$\Esubarg{\theta}{\sum_{i \in V} Y_i|\mathbf{Z}=\mathbf{1};G}$ is increasing with respect with its coefficient $\alpha_1$. If $\beta_0 = \beta_1$, the increasing functions
\begin{align*}
\alpha_0 &\mapsto \Esubarg{\theta}{\frac{1}{|V|}\sum_{i \in V} Y_i|\mathbf{Z}=\mathbf{0};G}\\
\alpha_1 &\mapsto \Esubarg{\theta}{\frac{1}{|V|}\sum_{i \in V} Y_i|\mathbf{Z}=\mathbf{1};G}
\end{align*}
are same. Therefore, 
$$\underbrace{\Esubarg{\theta}{\frac{1}{|V|}\sum_{i \in V} Y_i|\mathbf{Z}=\mathbf{1};G} - \Esubarg{\theta}{\frac{1}{|V|}\sum_{i \in V} Y_i|\mathbf{Z}=\mathbf{0};G}}_{\text{ATE}(\theta)} > 0$$
if and only if $\alpha_1 > \alpha_0$.

\section{Simulation Experiments}
We generate $K=100$ social networks $G^{(1)},\dots,G^{(k)}$, each of which contains $100$ users. Next, we randomly assign $Z_i^{(k)} \sim \text{Bernoulli}(1/2)$ i.i.d., and let $Y_i^{(k)}$ interact with each other in $G^{(k)}$ by Gibbs sampling on Ising/logistic model (\ref{eqn: Ising}). Finally, we got a simulation data set of a network A/B test for $\sum_{k=1}^K |V^{(k)}| = 10,000$ users. The methods presented in Section 2 are evaluated in three scenarios.


\subsection{Scenario I: Different Treatment Effect, Same Network Effect}
In the first scenario (Table \ref{tab: scenario 1}), the treatment coefficients are set as $\alpha_0=0.0$ and $\alpha_1 = 0.1$ such that
$$P_\theta(Y_i=+1| Z_i = 0, n(i) = \emptyset) = 0.50$$
$$P_\theta(Y_i=+1| Z_i = 1, n(i) = \emptyset) = 0.55$$
That means that user $i$, if assigned to group A, has probability of 0.50 to give positive response without the influence of neighbors; the treatment $B$ can increase the chance by 10\%. Other parameters $\beta_0 = \beta_1 = \gamma = \alpha_1/10$ means that the treatment effect is roughly equal to network and/or spill-over effects of 10 neighbors.

\begin{table}[h!]
\centering
\begin{tabular}{l l}
\hline
True Value & MPLE\\
\hline
$\alpha_0=0.00$ & $\hat{\alpha}_0=-0.00160$\\
$\alpha_1=0.10$ & $\hat{\alpha}_1=0.09372$\\
\hline
$\beta_0=0.01$ & $\hat{\beta}_0=0.00927$\\
$\beta_1=0.01$ & $\hat{\beta}_1=0.00443$\\
$\gamma=0.01$ & $\hat{\gamma} = 0.00304$\\
\hline
\end{tabular}
\caption{True parameter and MPLE in Scenario I where variations A and B have different treatment effect, but same network effect.}\label{tab: scenario 1}
\end{table}

\begin{figure}[h!]
\centering
\includegraphics[width=0.4\textwidth]{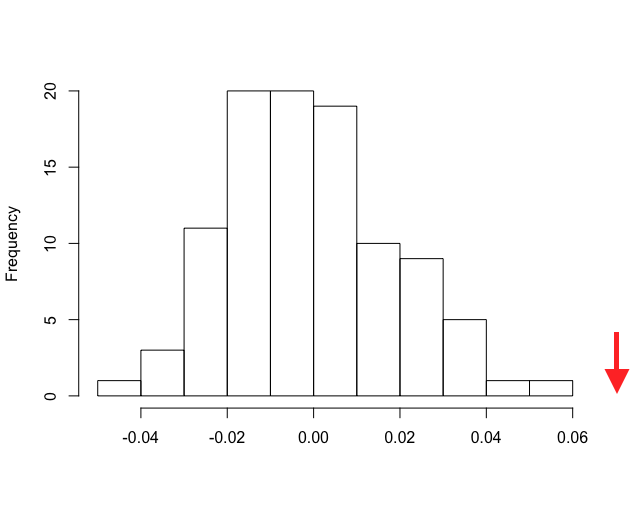}
\caption{Comparing $\hat{\alpha}_1 - \hat{\alpha}_0 = 0.095$ to the bootstrap distribution of $\tilde{\alpha}_1 - \tilde{\alpha}_0$ yields a p-value $<0.01$.} \label{fig: bootstrap I}
\end{figure}

Results show that $\hat{\alpha}_1 - \hat{\alpha}_0 = 0.095$ is quite close to the true value $\alpha_1 - \alpha_0 = 0.1$ Next, the bootstrapping method in Figure \ref{fig: bootstrap scheme} gives a p-value of $<0.01$, indicating the difference is significant (Figure \ref{fig: bootstrap I}).

\subsection{Scenario II: Same Treatment Effect, Same Network Effect}
\begin{figure}[h!]
\centering
\includegraphics[width=0.4\textwidth]{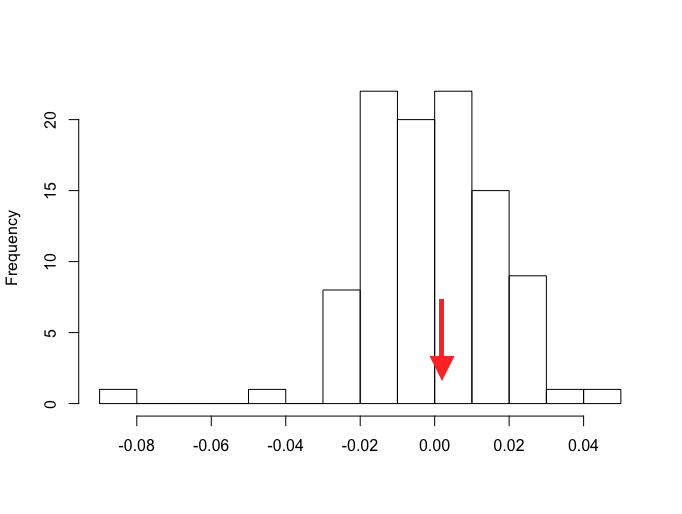}
\caption{Comparing $\hat{\alpha}_1 - \hat{\alpha}_0 = 0.004$ to the bootstrap distribution of $\tilde{\alpha}_1 - \tilde{\alpha}_0$ yields a p-value $=0.48$.} \label{fig: bootstrap II}
\end{figure}

The second scenario is an A/A test in which the treatment coefficients $\alpha_0= \alpha_1 = 0.05$ (Table \ref{tab: scenario 2}). The estimated value $\hat{\alpha}_1 - \hat{\alpha}_0 = 0.004$ is close to 0. And the the bootstrapping method in Figure \ref{fig: bootstrap scheme} gives a p-value of $=0.48$, indicating the difference is insignificant (Figure \ref{fig: bootstrap II}).
\begin{table}[h!]
\centering
\begin{tabular}{c c}
\hline
True Value & MPLE\\
\hline
$\alpha_0 = 0.05$ & $\hat{\alpha}_0 = 0.0448$\\
$\alpha_1 = 0.05$ & $\hat{\alpha}_1 = 0.0488$\\
\hline
$\beta_0 = 0.01$ & $\hat{\beta}_0 = 0.00761$\\
$\beta_1 = 0.01$ & $\hat{\beta}_1 = 0.00027$\\
$\gamma = 0.01$ & $\hat{\gamma} = 0.00607$\\
\hline
\end{tabular}
\caption{True parameter and MPLE in Scenario 2 where variations A and B have same treatment effect, and same network effect.}\label{tab: scenario 2}
\end{table}

\subsection{Scenario III: Same Treatment Effect, Different Network Effect}
\begin{figure}[h!]
\centering
\includegraphics[width=0.4\textwidth]{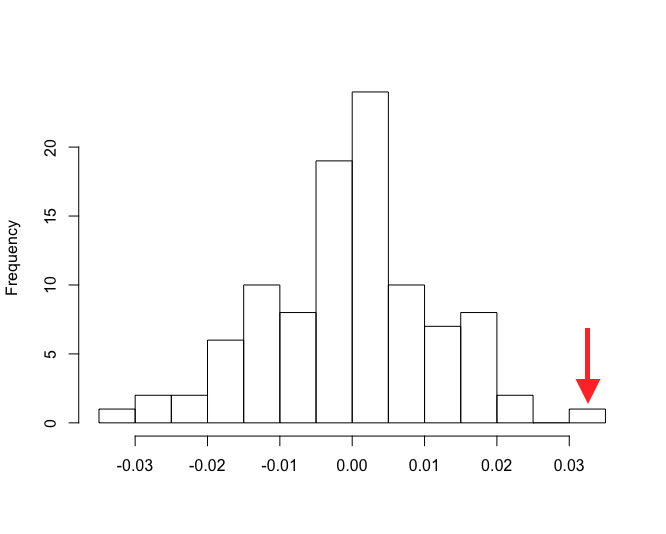}
\caption{Comparing $\hat{\beta}_1 - \hat{\beta}_0 = 0.004$ to the bootstrap distribution of $\tilde{\beta}_1 - \tilde{\beta}_0$ yields a p-value $=0.48$.} \label{fig: bootstrap III}
\end{figure}

In the third scenario, variation A and B have the same treatment effect on individual users ($\alpha_0 = \alpha_1 = 0.05$), but variation B has stronger network effect ($\beta_1=0.05$ v.s. $\beta_0 = 0.01$)and thus is more widespread in the social network. As shown in Table \ref{tab: scenario 3}, the estimated value $\hat{\beta}_1 - \hat{\beta}_0 = 0.039$ is close to the true value $0.04$. And the the bootstrapping method in Figure \ref{fig: bootstrap scheme} gives a p-value of $=0.03$, indicating the difference is significant (Figure \ref{fig: bootstrap III}).
\begin{table}[h!]
\centering
\begin{tabular}{c c}
\hline
True Value & MPLE\\
\hline
$\alpha_0 = 0.05$ & $\hat{\alpha}_0 = 0.0445$\\
$\alpha_1 = 0.05$ & $\hat{\alpha}_1 = 0.0485$\\
\hline
$\beta_0 = 0.01$ & $\hat{\beta}_1 = 0.00767$\\
$\beta_1 = 0.05$ & $\hat{\beta}_2 = 0.04626$\\
$\gamma = 0.01$ & $\hat{\gamma} = 0.01093$\\
\hline
\end{tabular}
\caption{True parameter and MPLE in Scenario III where variations A and B have same treatment effect, but different network effect.}\label{tab: scenario 3}
\end{table}

%
\bibliographystyle{abbrv}
\bibliography{ABtesting}  
%
%
\end{document}